\documentstyle[aps,prl,multicol,amssymb,epsf]{revtex}

\title{Bound Entanglement and Teleportation}

\author{N. Linden$^1$ and  S. Popescu$^{1,2}$}

\address{
$^1$Isaac Newton Institute for Mathematical
Sciences, Cambridge, CB3 0EH, UK\\%
$^2$BRIMS, Hewlett-Packard Laboratories, Stoke
Gifford, Bristol, BS12 6QZ, UK\\
}

\date{23 June 98}
\begin{document}
\draft
\maketitle
\begin{abstract}
Recently M. Horodecki, P. Horodecki and R. Horodecki have introduced a set of 
density matrices of two spin-1 particles  from which it is not possible to 
distill any maximally entangled states, even though the 
density matrices are entangled.  Thus these density matrices do not allow 
reliable teleportation.  However it might nevertheless be the case that these 
states can be used for teleportation, not reliably, but still with fidelity 
greater than that which may be 
achieved with a classical scheme.   
We show that, at least for some of these density 
matrices, teleportation cannot be achieved with better than classical fidelity. 
\end{abstract}


\begin{multicols}{2}


\newcommand\mathC{\mkern1mu\raise2.2pt\hbox{$\scriptscriptstyle|$}
                {\mkern-7mu\rm C}}		       
\newcommand{\mathR}{{\rm I\! R}}                

Non-locality, discovered by Bell more than thirty years ago, has recently shown 
itself to have many manifestations: teleportation\cite{BBCJPW}, 
distillability\cite{BBPSSW}, reduction  of communication 
complexity\cite{CleveBurhmann} etc. It is not clear yet what the relations 
between these different manifestations are \cite{Popescu94}.

In  two very interesting  papers \cite{Horodecki1,Horodecki2} a set of 
density matrices of two spin-1 particles  were introduced 
from which it 
is not possible to distill any states of which are  maximally entangled 
even though the density matrices  are entangled.  This is very surprising since 
in the case of spin-1/2 particles, any entangled density matrix is distillable. 
Distillability, however is just one aspect of non-locality.  Thus although this 
aspect of non-locality is not active, the Horodecki density matrices may 
actively manifest other forms of non-locality.  For example, bound entanglement 
may be pumped into a single pair of free entangled particles \cite{Horodecki3}. 
Here we investigate 
teleportation.

Since  one cannot distill pure singlets from Horodecki density matrices, these 
density 
matrices do not allow reliable teleportation. However it might nevertheless be 
the case that these states can be used for teleportation, not reliably, but 
still with fidelity greater than that which may be 
achieved with a classical scheme.  This is the general question we investigate 
here.

First let us consider the case of spin-1/2 particles. Since cannot distill 
states formally equivalent to spin-1/2 singlets from the 
Horodecki density matrices,
these density matrices cannot be used for reliable teleportation 
of spin-1/2.
Furthermore it is most probable that these density matrices cannot be used to 
teleport spin-1/2 states at all with better than classical fidelity. The reason 
is the following.  Suppose Alice produces locally a spin-1/2 singlet and 
teleports one of the spins to Bob. If Alice and Bob share a maximally entangled 
state of two spin-1 particles, then at then end of the process, this state is 
destroyed and it is replaced by a state which is equivalent to a maximally 
entangled state of two spin-1/2 particles (i.e. the state originally held by 
Alice).

If however Alice and Bob were to share a Horodecki density matrix, at the end 
of 
the process, the original Horodecki matrix is destroyed and now Alice and Bob 
share a pair of  particles whose state is equivalent to one of spin-1/2 
particles, but not a faithful copy of Alice's original state.  Presumably 
however, if the teleportation works better than any classical scheme, we expect 
that this state is still entangled.  This however cannot happen because from 
any 
entangled state of spin-1/2 particles one can distill singlets.  Thus the 
procedure would be tantamount to distilling singlets from the Horodecki 
matrices. We thus expect that
  Horodecki density matrices cannot be used to teleport spin-1/2  states with 
better than classical fidelity.

However the above discussion leaves open the question of whether spin-1 states 
can be teleported with better than classical fidelity using  Horodecki density 
matrices.  In particular the above argument does not rule this possibility for 
the following reason.  Suppose now  that Alice were to prepare a maximally 
entangled state of two spin-1 particles and teleport the state of one of them 
to 
Bob using a Horodecki pair.  At the end of the process the Horodecki matrix is 
again destroyed and Alice and Bob now share some state of two spin-1 particles.  
As before this state will not be a faithful copy of the original singlet 
prepared by Alice; however if the teleportation works better than any classical 
scheme, we expect that this state is still entangled.  But now there is the 
possibility that this state is a state of Horodecki-type {\em bound} 
entanglement which  does not allow distillation and thus leads to no 
contradiction of the Horodeckis' general arguments.  

Furthermore another argument which might give hope to the possibility that 
spin-1 states can be teleported better than the classical case while spin-1/2 
states cannot, is that one expects that the classical fidelity to be lower for 
spin-1 than spin-1/2. 
This is because it is more difficult to identify, using a measurement, a state 
which may 
be anywhere in a  3-dimensional Hilbert space than a state which may be 
anywhere 
in a 2-dimensional Hilbert space.

In this letter we show that, despite the arguments presented above, at least 
for 
some of the Horodecki density matrices, it is not 
possible to teleport spin-1 states with better than classical fidelity, thus 
confirming the remarkable nature of these density matrices. 

As is customary, we imagine that Alice and Bob each have one of the pair of 
particles described by the density matrix $\rho_a$.  Alice receives a particle 
in an unknown state $\phi$ and she performs a measurement of the pair of 
particles she has, and transmits some information to Bob.  The aim is to 
maximize the fidelity of transmission, averaged over all unknown states 
$\phi$.

To be explicit, we write the state $\phi$ as
\begin{eqnarray}
& &|\phi> = (c_5 + is_5 c_4)|1> + s_5s_4(c_3+is_3c_2)|2>\nonumber\\
& &\quad + 
s_5s_4s_3s_2e^{i\theta_1}|3>,
\end{eqnarray}
with respect to some basis $|1>,\ |2>,\ |3>$, and where $s_5=\sin\theta_5$, 
$c_5=\cos\theta_5$ etc. (for simplicity we have taken $\phi$ to be a vector 
rather than a ray). With this parametrisation, the $U(3)$ invariant 
measure is
\begin{eqnarray}
& & d\mu(\phi) = \nonumber\\
& &\quad{1\over \pi^3} \sin^4\theta_5 \sin^3\theta_4 \sin^2\theta_3
 \sin\theta_2\  d\theta_5  d\theta_4  d\theta_3  d\theta_2  d\theta_1. 
\end{eqnarray}

We first derive a general expression for the fidelity of transmission using an 
arbitrary density matrix $\rho$ shared between Alice and Bob.  A convenient 
parametrisation of $\rho$, the Schmidt representation, is
\begin{eqnarray}
\rho&=&{1\over 9} I\otimes I + {1\over 6}r_i \lambda_i\otimes I +
{1\over 6} I\otimes s_i\lambda_i\nonumber\\
 & &\qquad+ {1\over 4}t_{ij}\lambda_i\otimes\lambda_j, 
\end{eqnarray}
where $\lambda_i$ are the Gell-Mann matrices (see for example \cite{Cornwell})
which satisfy $Tr(\lambda_i)=0$ and $Tr(\lambda_i\lambda_j)=2\delta_{ij}$.
Thus $r_i = Tr(\rho \lambda_i\otimes I)$, $s_i = Tr(\rho I\otimes \lambda_i)$,
and $t_{ij} = Tr(\rho \lambda_i\otimes \lambda_j)$.

The fidelity of transmission is
\begin{eqnarray}
F=
\int d\mu(\phi)
& &\sum_{k=1}^9 p_k Tr_3(\rho_k P_\phi),
\end{eqnarray}
where
\begin{eqnarray}
p_k = Tr_{1,2,3}(P_k\otimes U_k)(P_\phi\otimes\rho)(P_k\otimes 
U_k^\dagger),
\end{eqnarray}
is the probability of the $k$'th outcome and 
\begin{eqnarray}
\rho_k = {1\over p_k}Tr_{1,2}(P_k\otimes U_k)(P_\phi\otimes\rho)(P_k\times 
U_k^\dagger),
\end{eqnarray}
is the output state.
\begin{eqnarray}
 P_k
 &=&{1\over 9} I\otimes I + {1\over 6}R^{(k)}_i \lambda_i\otimes I +
{1\over 6} I\otimes S^{(k)}_i\lambda_i\nonumber\\
 & &\quad+ {1\over 4}T^{(k)}_{ij}\lambda_i\otimes\lambda_j
\end{eqnarray} 
are the projection operators corresponding to the measurement Alice makes and 
$U_k$ are the unitary operators Bob performs which depend on which result 
Alice 
obtains.
\begin{eqnarray}
P_\phi={1\over 3} I + {1\over 2}\alpha_i \lambda_i
\end{eqnarray} 
is the projection operator of the unknown input state $\phi$.  The subscripts 
on the traces indicate the Hilbert space over which the trace is taken.

Now
\begin{eqnarray}
& &p_k\rho_k=\nonumber\\
& &\Big({1\over 27} + {1\over 18}r_qS^{(k)}_q + {1\over 18}\alpha_qR^{(k)}_q + 
{1\over 12}\alpha_p r_q T^{(k)}_{pq}\Big) I\nonumber\\
& &+ \Big({1\over 18}s_qO^{(k)}_{qj} + {1\over 12}t_{iq}S^{(k)}_iO^{(k)}_{qj} 
+
{1\over 12}\alpha_i R^{(k)}_i s_q O^{(k)}_{qj} \nonumber\\
& &\qquad + {1\over 8}\alpha_p 
T^{(k)}_{pi}t_{iq}  O^{(k)}_{qj}
\Big) \lambda_j
\end{eqnarray}
where the orthogonal matrix $O^{(k)}$ is that induced by  conjugation by the 
unitary matrix $U_k$:
\begin{eqnarray}
U^{(k)} x_j\lambda_j U^{(k)\dagger} = x_i O^{(k)}_{ij}\lambda_j.
\end{eqnarray}
Thus
\begin{eqnarray}
& &Tr_3(p_k\rho_k P_\phi) =\nonumber\\
& &\Big({1\over 27} + {1\over 18}r_qS^{(k)}_q + {1\over 18}\alpha_qR^{(k)}_q + 
{1\over 12}\alpha_p r_q T^{(k)}_{pq}\Big) \nonumber\\
& &+ \Big({1\over 18}s_qO^{(k)}_{qj} + {1\over 12}t_{iq}S^{(k)}_iO^{(k)}_{qj} 
+
{1\over 12}\alpha_i R^{(k)}_i s_q O^{(k)}_{qj} \nonumber\\
& &\qquad+ {1\over 8}\alpha_p 
T^{(k)}_{pi}t_{iq}  O^{(k)}_{qj}
\Big) \alpha_j.
\end{eqnarray}
We may now do integrals over $\alpha$ in the expression for the fidelity using
\begin{eqnarray}
\int d\alpha\ \alpha_i M_{ij}\alpha_j  &=& 
\int d\mu(\phi) <\phi|\lambda_i|\phi> 
M_{ij}<\phi|\lambda_j|\phi>\nonumber\\
 & =& {1\over 6} Tr(M)
\end{eqnarray}
and 
\begin{eqnarray}
 \int d\alpha\ \alpha_i  = 
\int d\mu(\phi) <\phi|\lambda_i|\phi> = 0.
\end{eqnarray}
Thus
\begin{eqnarray}
& & F=
\sum_k\Big({1\over 27} + {1\over 18}r_qS^{(k)}_q +
{1\over 72} R^{(k)}_p s_q O^{(k)}_{qp}\nonumber\\
& &\qquad + {1\over 48} T^{(k)}_{pi}t_{iq}  
O^{(k)}_{qp}
\Big).\label{fidelitysum}
\end{eqnarray}
We now put a bound on the fidelity by considering the maximum value of the 
summand.  Let us call $P^{max}$ (with Schmidt components $R^{max},\ S^{max}$, 
and $T^{max}$) the projection operator which maximises the 
summand in (\ref{fidelitysum}).  Without loss of generality we may take the 
orthogonal matrix $O$ to be the identity.  Thus
\begin{eqnarray}
& & F\leq
\Big({1\over 3} + {1\over 2}r_qS^{max}_q +
{1\over 8} R^{max}_q s_q 
 + {3\over 16} T^{max}_{pi}t_{ip}  
\Big).
\end{eqnarray}
Let us now denote by $\hat P^{max}$ the projection operator defined by
\begin{eqnarray}
\hat P^{max} = N P^{max} N,
\end{eqnarray}
where $N$ is the interchange operator which we define by its action on basis 
vectors:
\begin{eqnarray}
N e_i\otimes e_j = e_j \otimes e_i.
\end{eqnarray}
We may then use the fact that
\begin{eqnarray}
& &Tr(\rho \hat P^{max}) =\nonumber\\ 
& &\qquad {1\over 9} + {1\over 6} r_iS^{max}_i + {1\over 6} s_iR^{max}_i 
+ {1\over 4} t_{ij}T^{max}_{ji},
\end{eqnarray}
to rewrite the bound on the fidelity as
\begin{eqnarray}
& & F\leq
\Big({1\over 4} + {3\over 8}r_qS^{max}_q +
{3\over 4}Tr(\rho\hat P^{max})\Big).
\end{eqnarray}
We now consider the specific case of the matrices $\rho_a$ introduced in
\cite{Horodecki1,Horodecki2}:
\begin{eqnarray}
\rho_a={1 \over 8a + 1}
\left[ \begin{array}{ccccccccc}
          a &0&0&0&a&0&0&0& a   \\
           0&a&0&0&0&0&0&0&0     \\
           0&0&a&0&0&0&0&0&0     \\
           0&0&0&a&0&0&0&0&0     \\
          a &0&0&0&a&0&0&0& a     \\
           0&0&0&0&0&a&0&0&0     \\
           0&0&0&0&0&0&{1+a \over 2}&0&{\sqrt{1-a^2} \over 2}\\
           0&0&0&0&0&0&0&a&0     \\
          a &0&0&0&a&0&{\sqrt{1-a^2} \over 2}&0&{1+a \over 2}\\
       \end{array}
      \right ]. 
\label{rhoa}
\end{eqnarray}
We find that all the Schmidt components $r_q$ for this matrix are zero except 
$r_8$ 
which is
\begin{eqnarray}
r_8 = Tr(\rho_a \lambda_8\otimes I) = {2\over\sqrt{3}} \Big({a-1\over 8a 
+1}\Big).
\end{eqnarray}
Thus if we write 
\begin{eqnarray}
\tilde \rho_a = \rho_a + {1\over\sqrt{3}} \Big({a-1\over 8a 
+1}\Big)\lambda_8\otimes I,
\end{eqnarray}
then we may write the bound on the fidelity as
\begin{eqnarray}
& & F\leq
\Big({1\over 4}  +
{3\over 4}Tr(\tilde\rho_a\hat P^{max})\Big).\label{fidelitybound}
\end{eqnarray}

We now consider under what conditions the fidelity of teleportation can be 
greater than any classical procedure.  One particular classical scheme that 
Alice and Bob could use is as follows.  Firstly Alice simply  measures the 
unknown state $\phi$ using an arbitrary non-degenerate operator. Let us call 
the eigenvectors of this operator $v_1,\ v_2,\ v_3$ with associated 
eigenvalues $\mu_1,\ \mu_2,\ \mu_3$.  If Alice's outcome is 
$\mu_1$ she tells Bob to guess that the unknown state was $v_1$ and so on 
(this procedure may not be the optimal classical scheme, but we will not need 
this in what follows). The fidelity of this procedure is
\begin{eqnarray}
\sum_{i=1}^3 \int d\mu(\phi) |<v_i|\phi>|^4 = {1\over 2}.
\end{eqnarray}

Let us now return to the fidelity of teleportation.  The maximum value of 
the fidelity in (\ref{fidelitybound}) is obtained when we choose $ P^{max}$ so 
that $\hat 
P^{max}$ is the projector onto the maximum eigenvalue of $\tilde \rho_a$.  If 
this maximum eigenvalue is less that
${1\over 3}$, then the fidelity of teleportation (\ref{fidelitybound})  is 
less than
${1\over 2}$ and therefore the density matrix $\rho_a$ cannot be used to 
teleport better than the optimal classical scheme (which may have fidelity 
greater than ${1\over 2}$).

By direct calculation we find that for $a={\sqrt{3}\over 2}$ the eigenvalues 
of 
$\tilde \rho_a$ are
\begin{eqnarray}
& & {1\over 3} \Big( {2\sqrt{3} -1 \over 4\sqrt{3} +1}\Big), \nonumber\\
& &{83\over 1128} - {1\over 376}\sqrt{3}
-{1\over 376}(10588 - 5786 \sqrt{3}   )^{1\over 2},\nonumber\\ 
& &{83\over 1128} -{1\over 376}\sqrt{3}
+{1\over 376}(10588 - 5786 \sqrt{3}   )^{1\over 2},\nonumber\\
& &{7\over 141} -{3\over 94} \sqrt{3}  , \nonumber\\
& &{37\over 564} +{29\over 188} \sqrt{3},   \nonumber\\
& &{4\over 141} +{5\over 94} \sqrt{3};
\end{eqnarray}
the first eigenvalue occurs with multiplicity four.
All the above eigenvalues are less than ${1\over 3}$.  Thus we have shown that 
$\rho_{\sqrt{3}/ 2}$ cannot teleport a spin one state with fidelity better 
than classical.  We note that we have not limited ourselves to \lq\lq 
standard\rq\rq\ teleportation:  the projectors $P_k$ were not assumed to be 
maximally entangled.

Numerical evidence indicates that for $a$ roughly in the region
$4/5 < a < 1$, the maximum eigenvalue of $\tilde \rho_a$ is less than one 
third.  For small $a$, $\tilde \rho_a$ does have an eigenvalue larger than one 
third so the argument presented here is not conclusive in this case.

\bigskip

\end{multicols}

\end{document}